# Remarks on Bayesian Control Charts

**Amir Ahmadi-Javid†\* and Mohsen Ebadi†**

† Department of Industrial Engineering, Amirkabir University of Technology, Tehran, Iran

\* Corresponding author; email address: ahmadi_javid@aut.ac.ir

**Abstract.** There is a considerable amount of ongoing research on the use of Bayesian control charts for detecting a shift from a good quality distribution to a bad quality distribution in univariate and multivariate processes. It is widely claimed that Bayesian control charts are economically optimal; see, for example, Calabrese (1995) [Bayesian process control for attributes. *Management Science*, DOI: 10.1287/mnsc.41.4.637] and Makis (2008) [Multivariate Bayesian control chart. *Operations Research*, DOI: 10.1287/opre.1070.0495]. Some researchers also generalize the optimality of controls defined based on posterior probabilities to the class of partially observable Markov decision processes. This note points out that the existing Bayesian control charts cannot generally be optimal because many years ago an analytical counterexample was provided by Taylor (1965) [Markovian sequential replacement processes. *The Annals of Mathematical Statistics*, DOI: 10.1214/aoms/1177699796].

**Keywords:** Bayesian control charts, Economic design, Partially Observable Markov Decision Process (POMDP), Stochastic optimal control, Posterior probabilities.

# 1. Introduction

Bayesian control charts, originated by Girshick and Rubin (1952), are not new in the literature, and their economic design has received increasing attention over the last two decades (see Tagaras and Nikolaidis (2002), and Nikolaidis and Tagaras (2017), and references therein).

Girshick and Rubin (1952) discussed the optimality of Bayesian charts for the first time. They considered only a special case of a *discrete-time* production system that produces at discrete instants of time, and where 100% inspection is carried out (inspection costs are ignored). For this special setting, they studied the optimum quality control policy which specifies when to terminate production and put the machine in the repair shop in order to minimize the long-run expected average cost (see Ahmadi-Javid and Ebadi (2017) for important remarks on a class of optimal quality control problems with long-run expected average cost objective functions). They explicitly determined the following optimal policy (Tagaras, 1994):

"*Stop and repair at time t if and only if the posterior probability at time t that the process is in the bad state exceeds a control limit p.*"

Note that their proposed policy was initially presented in a different form. To see the equivalence to the above form, the readers are referred to the proof of Lemma 1 in page 116 of Girshick and Rubin (1952).

# 2. Discussion

Unfortunately, by generalizing the particular optimality result obtained by Girshick and Rubin (1952), several papers made misleading statements that imply the optimality of Bayesian control charts and the non-optimality of non-Bayesian control charts in general settings. In Table 1, a few of these statements are collected. This table provides a chain of citations starting from 1995.

Makis (2008) in the abstract of his paper explicitly claimed that he proved the optimality of a Bayesian control chart. However, he, and similarly Calabrese (1995) and Makis (2009), did not provide any proof and only cited the two papers Taylor (1965, 1967), where their statements have exactly the same wording. Let us now examine these Taylor's papers to make sure that they do not extend the optimality of Bayesian charts to more general settings, and that they only reassessed the results obtained by Girshick and Rubin (1952).

Table 1 A list of misleading statements regarding the optimality of Bayesian charts

| Paper | Journal | Statements |
| --- | --- | --- |
| Calabrese (1995) | *Management Science* | "**Taylor (1965, 1967)** has shown that non-Bayesian techniques are not optimal …" (page 637. line 16) |
| | | "A Bayes statistic together with a simple control limit policy is shown to be an economically optimal method of process control" (page 638, line 9) |
| Makis (2008) | *Operations Research* | "It is well known that these traditional, non-Bayesian process control techniques are not optimal "( Abstract: page 795, line 4) |
| | | "Under standard operating and cost assumptions, [in this paper] **it is proved that a** [Bayesian] **control limit policy is optimal**, and an algorithm is presented to find the optimal control limit and the minimum average cost."(Abstract: page 487, line 8) |
| | | "**Taylor (1965, 1967)** has shown that non-Bayesian techniques are not optimal …" (page 488, line 4) |
| Makis (2009) | *European Journal of Operational Research* | "It is well known that this traditional non-Bayesian approach to a control chart design is not optimal" (Abstract: page 487, line 3) |
| | | "**Taylor (1965, 1967)** has shown that non-Bayesian techniques are not optimal…" (page 796, line 9) |
| Wang, & Lee (2015) | *Operations Research* | "**Calabrese (1995) proved** that, when the sampling interval and the sample size are both fixed, the [Bayesian] control-limit policy is optimal for finite-horizon problems. **Makis (2008, 2009)**, assuming a binary state space, **showed** that the control-limit policy is optimal for multivariate control charts in the finite and infinite-horizon cases" (page 1, line 46). |

Taylor (1965) studied a general control problem, called a sequential replacement process, which deals with a dynamic system that is observed periodically and classified into one of a number of possible states; and after each observation, one of possible decisions is made. A sequential replacement process is a control process with an additional special action, called replacement, which instantaneously returns the system to some initial state. The paper first proves the existence of an optimal stationary non-randomized rule, and then presents a method to determine the optimal policy under two popular effectiveness measures: the expected total discounted cost and the long-run excepted average cost. The paper also shows that the optimal rule proposed by Girshick and Rubin (1952) for "100% inspection" case can be covered by their results. On the other hand, using a counterexample, the paper shows the non-optimality of the other rule that Girshick and Rubin (1952) claimed to be optimal for the "non-100% inspection" case. Taylor (1967) discussed no optimality result and only proposed a method to approximately determine the optimal control limit $p$ of the rule proposed by Girshick and Rubin (1952) for "100% inspection" case whenever the bad state slightly deviates from the good state.

Hence, Taylor (1965, 1967) did not provide any new optimality results for Bayesian charts. In fact, Taylor (1965) proved the fallacy of the Girshick and Rubin's claim on the optimality of Bayesian charts in a general case.

We just received responses of two experts in the field of Bayesian control charts on the above facts. They again expressed a convinced belief in the optimality of Bayesian charts and generalized it to the class of partially observable Markov decision processes. To prove their claims, they just provided another three references Bertsekas and Shreve (1996), Davis (1993), and Smallwood and Sondik (1973), in which we could not find any proof. Actually, they stated:

> "*As proved e.g. in Bertsekas and Shreve: Stochastic Optimal Control (see also Davis: Markov Models and Optimization), Bayesian control is indeed optimal for partially observable stochastic processes, where the posterior probability is a sufficient statistic for optimal dynamic decision-making.*"

> "*I personally do not consider this proof necessary in these papers because it is a well-known result in the theory of partially observable Markov decision processes (POMDP), see Smallwood and Sondik (1973).*"

These researchers who responded to our manuscript did not provide any reference that proves their claims. Moreover, they did not show that the counterexample provided by Taylor (1965) was incorrect.

## 3. Conclusion

An important open area is to characterize the class of optimal quality-control charts for the discrete-time or continuous-time production systems considered in Girshick and Rubin (1952), Calabrese (1995), Makis (2008), and many other papers with slight differences. It seems that the current threshold-based Bayesian control charts cannot be optimal. The reason is that Taylor (1965) showed that the Bayesian charts were not optimal even for the discrete-time production setting with non-100% inspection, first considered by Girshick and Rubin (1952). Moreover, characterizing optimal controls for the general class of partially observable Markov decision processes that cover the above quality-control problems remains as another future research direction.